# Elements of a Theory of Simulation

Steen Rasmussen[1,2] and Christopher L. Barrett[1,2]

[1] Los Alamos National Laboratory, Los Alamos NM 87545, USA
[2] Santa Fe Institute, 1399 Hyde Park Road, Santa Fe NM 87501, USA

**Abstract.** Artificial Life and the more general area of Complex Systems does not have a unified theoretical framework although most theoretical work in these areas is based on simulation. This is primarily due to an insufficient representational power of the classical mathematical frameworks for the description of discrete dynamical systems of interacting objects with often complex internal states.

Unlike computation or the numerical analysis of differential equations, simulation does not have a well established conceptual and mathematical foundation. Simulation is an arguable unique union of modeling and computation. However, simulation also qualifies as a separate species of system representation with its own motivations, characteristics, and implications. This work outlines how simulation can be rooted in mathematics and shows which properties some of the elements of such a mathematical framework has.

The properties of simulation are described and analyzed in terms of properties of dynamical systems. It is shown how and why a simulation produces emergent behavior and why the analysis of the dynamics of the system being simulated always is an analysis of emergent phenomena. Indeed, the single fundamental class of properties of the natural world that simulation will open to new understanding, is that which occurs only in the dynamics produced by the interactions of the components of complex systems. Simulation offers a synthetic, formal framework for the experimental mathematics of representation and analysis of complex dynamical systems.

A notion of a universal simulator and the definition of simulatability is proposed. This allows a description of conditions under which simulations can distribute update functions over system components, thereby determining simulatability. The connection between the notion of simulatability and the notion of computability is defined and the concepts are distinguished. The basis of practical detection methods for determining effectively non-simulatable systems in practice is presented.

The conceptual framework is illustrated through examples from molecular self-assembly end engineering.

# 1 Introduction

It is typical in both science and engineering to be interested in properties or causal details of a phenomena for which we do not have an adequate, explicit model. For instance this is the case when we are investigating the complicated phenomena of life. In other less complicated situations a model of the phenomena may not yet be derived; indeed, the derivation may be the very question under investigation. Also, analytic solutions may not be tractable due to inherent system complexity.

In these situations it is increasingly common to resort to *synthetic* methods using computer simulation. In a certain sense then, the comments that follow may seem as merely common sense. However they lead, we believe, to genuinely important considerations that can be briefly summarized in the following paragraph.

The essence of synthetic methods is that a simulation is a mechanism which interacts many state transition models of individual subsystems (i.e., system components) and thereby *generates* system dynamical phenomena. There is nothing inherently explanatory in this stage of investigation, it is essentially representational. The issues are that the system dynamical representation is *implicit* and *constructive*: The relations that constitute the properties of interest are nowhere explicitly encoded in the simulated component subsystems, but rather *emerge* and become accessible to *observation* as a result of the collective effects of computed interactions among these subsystems. This implies that certain assumptions have been made concerning the status of the capabilities of simulation as a species of representation in relation to computability, observability, numerical stability, and many other issues.

In addition to these important and formalizable issues are the need to establish an elementary and general concept of the generative concept of emergence. That is to separate the dynamics of the simulation mechanism, which is itself an iterated system, from the simulated system, which itself is represented in the framework provided by the simulation system. Thus, there is a need to relate the concept of emergence and the concept of simulation. Some clarification of at least the basic form of these issues is the aim of this paper.

Consistent with what is summarized in the above paragraph, the general research, or indeed any application-specific, program of investigating or developing control strategies for complex dynamical phenomena using simulation is in four basic parts: (i) We must always be aware that a simulation is generating dynamical phenomena at a level which is *higher* than the level from which the elemental interactions are described. If we are to exploit simulation it is necessary to understand what this capability to produce hierarchies of emergent relations implies. (ii) We must have methods with which to identify the elements of the underlying system that create the phenomena of interest. (iii) It is then necessary to formulate models of the important underlying subsystems (those that define the elemental subsystems and the element-element or object-object interactions). (iv) Finally, we must create the framework in which the simulation of the subsystems in interaction is composed, and embody the system represen-tation in that framework so that the phenomena of interest can be generated and analyzed.

Part (i) does not seem to have been treated in general terms of dynamics, which is odd, since it is the foundation of all simulation-based work in many disciplines. Part (ii) is problem specific, but general principles do exist such as those mentioned above that, in one form or another, occupy the attention of systems science. Part (iii) involves how an appropriate "level of aggregation" and useful "collective coordinates" can be chosen. These are not simple questions. Among other things, aggregation depends on which global system properties are of interest, how the component representations can be made as parsimoniously as possible, and what can be observed about the system behavior. Parts (iii) and (iv) can be combined to form a single, broader, question: "Given a system composed of many interacting subsystems, how does one formulate models of the subsystems and cause them to interact in a simulation environment that will generate appropriate global system dynamics?" Or the Artificial Life variant: "How can we generate life-like behaviors using low-level, local rules?" Over the years, a variety of proposals has been given in an attempt to answer this question, depending on the characteristics of the system under investigation, the kind of system properties of interest - as well as taste.

In large complex systems the system, in the sense of the generator of the dynamics of interest, is implicitly presumed to exist. For example, we presume that a solution of polymers with hydrophilic heads and hydrophobic heads in water is indeed a system. Furthermore, we assume that it is a dynamical system in the sense that the state of the system at time, $t$, and some "state transition rules" completely determine the state of the system at time, $t + \Delta t$. That is to say that the system has a "model". However, we doubt the possibility of writing down an explicit analytical expression of these dynamics in terms of all of the relevant state variables and parameters and, therefore, the model of that system. Moreover, we doubt the tractability of the solution of such a model even if one could be somehow defined.

However, we do not discount the possibility of modeling each of the relevant elemental subsystems in isolation. By "elemental" subsystems we mean as monomers and solvent molecules [11], or perhaps as vehicles, roadway segments, signals, and travel goals in a traffic system [3] [9]. We can imagine various specialist-practitioners to be able to define or, at least, hypothesize possible relevant ental subsystems and characterize local interaction rules. We can imagine modeling these individual subsystems and interaction rules well enough to define the state transition and interaction possibilities of a single class of subsystems that could be inherited by every instance of that class.

Given an object-entity-subsystem perspective, interactions can in general be viewed as *discrete events* among the subsystems undergoing local state changes and communicating these state changes to its neighbors in some space. That is to say that the interactions can be viewed as calculable by means of discrete

---

[3] As in TRANSIMS, an ongoing, large scale Transportation ANalysis and SIMulation System project at the Los Alamos National Laboratory.

event, object-oriented, simulation of collections of subsystems.

The concept of an *event driven simulation* contains the most general updating scheme for a simulation, since an event can either be externally or internally generated, and, as a special case, an event can also be *a time step*. Thus, a time stepped simulation is a special case of an event driven simulation, namely a discrete event simulation where the update is driven by the event that a clock-entity object produces as its internal state and transmits to all other objects. It is perhaps more accurate to say that any time stepped method can be simulated in some event driven method. It is, however, not the purpose of this paper to develop new update schemes or to review the extensive literature in this area, (e.g. Jefferson[6] and Lubachevsky[7]). We only mention the issue here to clarify our use of language and basic concepts as well as to set the stage for what we do intend to investigate. These issues will include the most general issues of the concept of object state update and the coordination, the scheduling, of updates of interacting object-subsystem.

The properties and consequences of a generalization of the simulation scheduling problem, together with the notion of hierarchies of emergent dynamical relations, seem to us to form the elementary foundations of simulation rooted in dynamics.

We have for the current presentation mostly focused on discrete space and discrete time systems (mappings), which, for the most part, are defined through interacting objects with some (minimal) internal state complexity. We have chosen to do so because the proposed framework is natural for such systems, but it should be noted that continuous space, continuous time, dynamical systems, equally well can be treated in the given framework. Moreover, discrete space, discrete time systems do not have any other general formal framework within which the dynamical properties can be generated and analyzed.

## 2 Simulation

This paper describes simulation as an *iterated mapping* of a (usually large and complicated) system. The simulated system is usually decomposed to a level where subsystems or system components are individually defined as encapsulated objects that calculate and communicate internal state. The simulation is an iterative system in which the simulated system is represented and its dynamics calculated. Thus the simulation and the simulated system are both dynamical systems and the interplay between the dynamics of the coordination of the simulation updates and the dynamics calculated in the time series of system states are essential issues.

In the above paragraph, we have distinguished four "systems" that comprise a simulation. We assume the existence of some $\Sigma_R$, a real or natural system in the world that we are interested in, $\Sigma_{(S_i \in M)}$, models of subsystems $S_i$ of this system and rules that define interactions among the subsystems, $\Sigma_S$, a simulation of $\Sigma_R$ involving $\Sigma_{(S_i \in M)}$ and some update functional $U$, and finally, $\Sigma_C$, a formal (and equivalent physical) computing machine on which $\Sigma_S$ is implemented.

*Definition*: The *objects* (elements or subsystems) in a simulation are defined as

$$S_i = S_i(f_i, I_{ij}, x_i, t_i), i \text{ and } j = 1, ..., n, \qquad (1)$$

where $f_i$ is the representation of the dynamics of the $i$th object and where $I_{ij,j=1,...,n}$ is the $i$th object's interaction rules with other objects $j$. Interaction and dynamics operate on $x_i$, the state of the $i$th object. $t_i$ is the local object time coordinate.

*Definition of* $\Sigma_{(S_i \in M)}$: $S_i$ is an element in the system $\Sigma_{(S_i \in M)}$; that is, $S_i$ is a *model* of the $i$th element of the set of modeled system elements in $M$, $i = 1, ..., n$. Thus, the algorithmic part of $S_i$ is equivalent to $f_i$ and $I_{ij}$.

*Definition of* $U$: An object *update functional* $U$ is the state transition

$$S_i(t_i + \Delta_i) \leftarrow S_i(t_i), i = 1, ..., n, \qquad (2)$$

or

$$S_i(t_i + \Delta_i) = U(S_i(t_i)), i = 1, ..., n, \qquad (3)$$

where $U$, the update functional, defines, organizes and executes the formal iterative procedure that prescribes the state transition.

*Definition of* $\Sigma_S$: A *simulation* is the *iteration* of object updates over the entire set of objects

$$\{S_i(t+1)\} \leftarrow \{S_i(t)\}, i = 1, ..., n. \qquad (4)$$

or

$$\{S_i(t+1)\} = U(\{S_i(t)\}), i = 1, ..., n. \qquad (5)$$

A valid update functional $U$ also needs to be able to time align all objects at regular intervals or at a given time, perhaps at each update. Note that $f_i$ together with $I_{ij}$, $x_i$ and $U$ *implicitly* defines the dynamical properties of the system. $U$ can be viewed as the "active" part of $\Sigma_S$ where as $f_i$, $I_{ij}$ and $x_i$ can be viewed as the "passive" parts of $\Sigma_S$.

Thus, the iteration of the dynamics of $\Sigma_{S_i \in M}$ using $U$ constitutes a formalization of $\Sigma_S$, the simulation system.

*Definition of* $\Sigma_C$: The formal, or equivalent physical implementation, of the mechanisms of the iteration procedure that prescribe the interactions and consequent object state transitions and their storage. $\Sigma_C$ is normally a physically and conceptually digital computer of some kind.

## 3 Emergence

Having defined $n$ objects or structures $S_i^1 \in \Sigma_M$ and an update functional $U$ at some level of description, say $L^1$, we now also introduce an observational function $O^1$ by which the objects can be inspected. Iterating $\Sigma_M$ using $U$ a new structure $S^2$ may be produced over time

$$S^2 \leftarrow U\{S_i^1(f_i, I_{ij}^1, x_i, t_i), O^1\}, i \text{ and } j = 1, ..., n. \tag{6}$$

This is what we call a *second order structure* occuring at level $L^2$. This new structure may be subjected to a possible new kind of observer $O^2$.

*Definition:* We define that a property $P$ is *emergent* iff

$$P \in O^2(S^2), \text{ but } P \notin O^1(S_i^1). \tag{7}$$

In this sense emergence depends essentially on the observer in use which may be *internal* or *external*. It should be noted that the generated, emergent properties may be *computable* or *non-computable*. For a comprehensive discussion of emergence we refer to [1].

This process can be iterated in a *cumulative*, not necessarily a *recursive*, way to form *higher order emergent structures* or *hyperstructures* of e.g. order $N$:

$$S^N \leftarrow U(S_i^{N-1}, O^{N-1}, S_k^{N-2}, O^{N-2}, ...) \tag{8}$$

Note that the definition of an observation function is no more - or just as - arbitrary as the definition of the objects and their interactions.

Examples of emergent properties could for instance be the dynamical properties of a polymer in solution or the properties of a congestion in a traffic system. The polymer as well as the congestion can be viewed as $S^2$ structures. A lower level $L^1$ description of the interactions will in the polymer example mean to describe the monomer-monomer interactions together with the monomer-solvent molecule interactions ($S_i^1$-$S_j^1$ interactions). In the example of traffic congestion it means to describe the vehicle-vehicle interactions together with the vehicle-roadway and -signal interactions (again $S_i^1$-$S_j^1$ interactions).

In these examples the $S_i^1$ interactions generate the $S^2$ phenomena, but the $S^2$ structures also have a *downward causal effect* on the $S_i^1$ structures. The polymer restricts the dynamics of the monomers, that it is made up of. The jam does the same, it also restricts the dynamics of the vehicles it is made up of.

However, emergent properties, as defined above, may not always have a downward causal effect. For example, the joint distribution of heads and tails generated from two independent coin flips is an emergent property of the system, but the distribution does obviously not have any influence on the dynamics of the coins.

A central question to ask here is: What is the *minimal* (or *critical*) *object complexity* needed to generate an emergent property of a given order in $\Sigma_S$? Complexity here refers to computational complexity, which may be defined through the time (or number of steps) or the capacity (memory), which at a minimum, is needed to generate the particular property [10].

## 4 Simulation and Emergence

It is in the general case very difficult, and perhaps in some cases even impossible, to come up with a direct, a priori description (a model) of the dynamics of the phenomena $S^2$ of interest in systems consisting of many, interacting elements with some internal complexity. In general it may, however, be possible to identify the level, say $L^1$, from which the phenomena of interest *emerges* and where it in a direct way is possible to describe the interactions or the dynamics of the elements or objects that generate $S^2$.

If we assume that a formal description of the object-object interactions is possible at level $L^1$ and that some observation mechanism $O^2$ exists so that properties of $S^2$ can be detected and their dynamics followed, then the situation is the following at level $L^1$: Explicit models $S_i^1$ exist to describe the dynamics of and the interactions between the $n$ objects where the object's states depend on each other. However, a global state dynamics function $F^1$ may only *implicitly* exist at level $L^1$. $F^1$ is the global function that describes the system wide state changes caused by the object-object interactions described by the set of local $f_i$s and $I_{ij}$s. The total system state $\chi^1(t)$ at level $L^1$ can be obtained via appropriate observational functions $O^1$ successively applied to each of the objects. Thus,

$$\chi^1(t) = \{x_1^1(t), ..., x_n^1(t)\}. \tag{9}$$

The state dynamics function $F^1$ is therefore always at least implicitly given at level $L^1$, since $\chi$ can be computed at any time. Thus the description of the $L^1$ dynamics is in principle known on the form

$$\chi^1(t+1) \leftarrow F^1(\chi^1(t)). \tag{10}$$

To actually *produce* the dynamics some update functional $U$ is needed which is able to organize the update of the interacting set of objects in a consistent way. Assuming that some update functional $U$ exists we have

$$\{S_1^1(t+1), ..., S_n^1(t+1)\} = U(\{S_1^1(t), ..., S_n^1(t)\}) \tag{11}$$

or

$$\chi^1(t+1) = U(F^1(\chi^1(t))). \tag{12}$$

Thereby the dynamics of system $\Sigma_S$ can be *generated*.

From the above it is clear that whenever it is possible to define an update functional $U$ that can organize the interactions of the objects defined at level $L^1$ through the set of models $M$, then the $L^2$ phenomenon of interest $S^2$ *emerges* and can be followed, applying the observation function $O^2$. Note that this is possible even without knowing $F^1$ explicitly. Thus, a recursive application of $U$ to the objects *generates* $S^2$ and the dynamics of $S^2$ (which is a property $P^2$ of $S^2$) can then be followed by a recursive application of $O^2$.

The central point is that *a simulation is a representational mechanism that is distinguished by its capacity to generate relations that are not explicitly encoded.*

Recall that $S^2$ for instance could be a polymer described through monomer-monomer and monomer-solvent interactions. In that situation $P^2$ could be the polymer elasticity. $S^2$ could also be a traffic jam described through vehicle-vehicle and vehicle-roadway interactions and then $P^2$ could for instance be the lifetime of a jam.

Thus, we have
$$S^2 \leftarrow U(\{S_1^1, ..., S_n^1\}) \tag{13}$$
and
$$P^2 = O^2(S^2). \tag{14}$$
Note that $S^2$ in (13) is defined through the implicit (emergent) relations that are generated between the objects on the left hand side of (11).

Recall that we in principle would like to be able to follow the state dynamics of $\Sigma_R$ through some $\Sigma_M$ at level $L^2$ in a direct way. But this requires that the state variables $\{x_1^2(t), ..., x_m^2\} = \chi^2(t)$ together with the state dynamics function $F^2$ at level $L^2$ were known explicitly, thus that we could write

$$\chi^2(t+1) \leftarrow F^2(\chi^2(t)), \tag{15}$$

which expresses that the state dynamics can be derived from the current state by applying some $F^2$. Knowing $F^2$ would in principle also enable some update functional $U^2$ to produce the dynamics

$$\chi^2(t+1) = U^2(F^2(\chi^2(t))). \tag{16}$$

Since we assume that the system cannot *a priori* be described at level $L^2$, but that it can be described at level $L^1$ the dynamics at level $L^2$ can be *generated* by *simulating* the interactions of the objects $S_1^1, ..., S_n^1$ at level $L^1$. In other words: By simulating the interactions of $S_i^1$ at level $L^1$ the phenomena and relations of interest at level $L^2$ will *emerge*.

Note that simulation is a direct generative way to obtain knowledge of this kind of non-explicitly encoded (dynamical) relations and phenomena. *Simulation is therefore a natural method to study emergence.* The non-explicitly encoded relations may later explicitly be modeled in a closed form, but that is irrelevant.

In fact, science if full of descriptions of systems where we have both an $L^1$ and an $L^2$ description. Classical examples include the Statistical Mechanical ($L^1$) versus the Thermodynamical ($L^2$) description of matter as well as the Lattice Gas Automata for fluid particle dynamics ($L^1$) versus Navier Stokes equations for macroscopic fluid dynamics ($L^2$).

## 5  Simulatability

A major remaining issue concerning simulation is: Under which conditions does an update functional exist for a large number of different, interacting objects? It is obvious that the object interactions can be rather involved and thus difficult to "untangle" so that the objects actually can be updated.

Let $\varphi(S_q^1, ..., S_s^1)$ be a hierarchically distinct representation of a subset of the interacting objects $S_q^1, ..., S_s^1$. Thus, $\varphi(S_q^1, ..., S_s^1)$ defines a sub-aggregation (an aggregated model) of some of the objects.

*Definition*: If

$$U(\{S_1^1, ..., S_n^1\}) = \{U(S_i^1), U(S_j^1), ..., U(\varphi(S_q^1, ..., S_s^1)), U(S_l^1), ..., U(S_p^1)\} \quad (17)$$

for some order of the objects, then the update $U$ is *distributable* over the decomposition $\Sigma_M$ of the system and each object and object aggregation can be updated independently of each other.

Note that if $U$ is operating in a sequential manner the list on the right hand side of (17) is ordered. Thus the sequence in which the objects are updated matters. If $U$ is operating in a parallel manner the order in which the objects are updated does not matter.

   A simple example of a *subaggregation* in a simulation is a particle collision in a lattice gas automata [4] [5]. As long as the fluid particles do not interact they are updated independently of each other - they just propagate along the lattice. But when they collide they are aggregated and the individual particle velocities after the collision are given by a collision table which takes the incoming, colliding particle velocities as arguments.

The nature of the update functional $U$ has a significant influence on the dynamics. The same model decomposition $M$ will in general generate different dynamical properties if different update functionals $U$ are applied. For instance, the elementary 1D, radius one cellular automata with rule 58 (00111010) will exhibit very different dynamics using a parallel or a random update respectively. Obviously, both the random and the parallel update *distributes* over any of the elementary rules on the 1D lattice.

   It is clear that the representation of the objects and their interactions, $M$ (the models of interactions at $L^1$) is crucial for whether a given update $U$ can distribute or not. It is assumed that each object, given the $(M, U)$ pair, individually can be updated when supplied with the appropriate state information from its communicating objects.

Thus, a system is *simulatable* iff there exists a pair $(M, U)$ such that $U$ *distributes*

over $S_i \in M$.

The above follows directly from our assumptions and definitions. Since it is assumed that each object, or sub-aggregation of objects, given the $(M, U)$ pair, can be updated individually and that problems can only occur due to the order and the organization of the object-object interactions. Since each of the objects and/or sub-aggregations can be updated independently in a system where the $(M, U)$ pair allows the update to distribute the above follows.

A direct consequence of this is that if no distributable $U$ exists for some sub-aggregation of the objects $M$ which allows the update to distribute, then the system is non-simulatable. Conversely, if no $M$ exists so that a given $U$ can distribute then the system is also non-simulatable.

A situation may occur where the smallest sub-aggregation which can be updated independently is the system itself. In this situation the sub-aggregation defines a *model* for the whole system at level $L^1$.

An example of a *non-simulatable* systems consider a model polymer defined on a 2D lattice [11]. Assume that the polymer is embedded in some solvent (heat bath) and that we would like to update each of the monomers in parallel. To perform the update and thus generate a possible new (lattice) position for each of the monomers in the polymer the polymer should not break and it should respect its excluded volume. That means that each monomer requires information about every other monomer in the polymer to be able to resolve possible conflicts due to the no-break and the excluded volume constraints. Thus, the minimal complexity of the model $M$ depends on the polymer length! Only allowing, say $k$ steps, in the update cycle (in the model of monomer-monomer and solvent-monomer interactions) implies that polymers above a certain (finite) length are non-simulatable, because the update does not distribute over the objects, since the objects cannot be updated independently of each other after they have communicated with each other.

However, defining a update scheduling such that alternating monomers are updated in every first and every second part of the update cycle, polymers of any length can be simulated. The monomer models $M$ becomes much simpler using such a two step parallel scheduling instead of using a strict parallel update [11].

## 6  A Universal Simulator

We define a universal simulator $US$ as a machine that is able to resolve all causal dependencies among the objects $S_i^1$. Thus, a $US$ can determine whether the system given the $(M, U)$ pair is simulatable or not. Further it can give an appropriate order of updating for the objects if it is simulatable and detect where the problems are if the system is non-simulatable. Thus, the *scheduling problem* lives in the $US$. Since the causal dependencies is being done on-line it may be of particular, practical interest for event driven simulations.

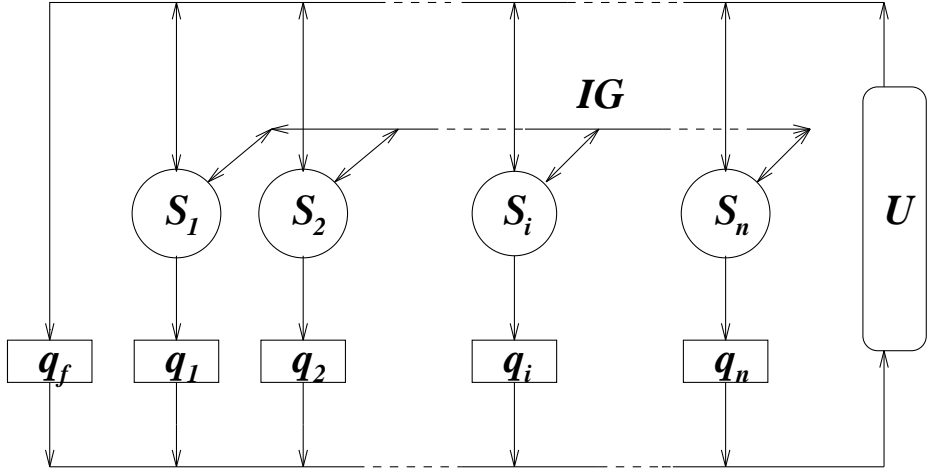

**Fig. 1.** *A universal simulator. The objects are denoted $S_i^1$, the $q_i$'s denote counters associated with the objects, $\mathbf{q} = (q_1, ..., q_n)$, $Q = \sum_i (q_i)$ (the sum of all the individual object counters). $q_f$ counts the failed update trails, the update functional is denoted $U$ and the interaction graph is denoted $IG$.*

The structure of a $US$ is defined in figure 6.

To perform a complete update of $\Sigma_S$ each of the objects $S_i^1$ need to be updated at least once and all need to be aligned in time. Assume that we are at *time* 1. The dynamics of the machine is as follows: Start by attempting to update $S_1^1$. If $S_1^1$ does not depend on the state of any other object at *time* 1 it is updated and its associated counter $q_1$ is incremented by one. If, however, $S_1^1$ has dependencies (depends of the state of one or more of the other objects at *time* 1) the object $S_1^1$ is exited and $q_f$ is incremented and the next object $S_2^1$ is attempted to be updated. If also $S_2^1$ has dependencies $q_f$ is again incremented and object 3 is attempted to be updated and so forth until an updatable object, say $S_i^1$ is found.

Without any loss of generality assume that all objects have internal dynamics on the same time scale (no objects need to be updated with any smaller time resolution than any other) and that $U$ is a discrete time update.

There are $n$ ways to pick the first object to update and each of these objects have at most $(n-1)$ objects they can depend on. Thus an upper bound on the number of operations needed to find the first updatable object is $n(n-1)$, given that it exists. The second updatable object does at most require $(n-1)$ operations to find and it takes at most $(n-1)$ operations to check whether any of the other objects influence it. Thus, an upper bound on the number of operations it takes to update the whole system once is given by

$$\sum_{i=0}^{n-1}(n-1)(n-i). \tag{18}$$

As an upper bound, the universal simulator can therefore determine whether a system is simulatable or not in at most $\sum_{i=0}^{n-1}(n-1)(n-i)$ operations.

As the update steps through the $US$-algorithm that sorts out the object inter-dependencies it simultaneously defines the *object update dependency Jacobian*.

$$D\mathbf{q} = (\frac{\partial q_i}{\partial q_j})_{n \times n} \equiv (\frac{\Delta q_i}{\Delta q_j})_{n \times n} \tag{19}$$

where the derivative $\partial q_i/\partial q_j$ expresses how many updates of object $j$ are necessary to update object $i$ once. Thus $\Delta q_i/\Delta q_j = 0$ indicates that the update of object $i$ is independent of object $j$ and $\Delta q_i/\Delta q_j = 1$ (or any natural number larger than one) tells that the update of $i$ needs the state of object $j$ at current time before it can be updated. As a special case $\Delta q_i/\Delta q_j \equiv 1$ for $i = j$.

A system $\Sigma_S$ is *simulatable* if $D\mathbf{q}$ is a diagonal matrix (unit matrix). This follows directly from the definition of the object update dependency Jacobian.

If $D\mathbf{q}$ contains sub-matrices on the diagonal and $\Sigma_S$ is found non-simulatable it is an indication that $\Sigma_S$ could become simulatable by the construction of sub-aggregations including the objects contained within each of the sub-matrices. Recall the definition of simulatability in section 5.

In the case of an upper diagonal object update dependency Jacobian a re-organization of the ordering of the updates can make a diagonal matrix.

Note that a parallel updatable system as well as a strictly sequential updatable system have the same structure in their update dependency matrices. If $D_i\mathbf{q}$ is a diagonal matrix it only insures that the system is simulatable and it gives a causal order of the object updates. Other orderings may exist.

Since $\partial Q/\partial q_i$, $Q = \sum_i (q_i)$, defines the relative computational load of the $i$th object the problem of *load balancing* also naturally lives in the $US$.

Also note that since the matrix $D\mathbf{q}$ is purely emperical, that is it evolves during the course of the simulation, the control problem associated with coordinating the simulation update is formalizable in terms of the trajectories of $D\mathbf{q}$ as a function of the update actions.

## 7 Achievable Simulatability

Suppose we are given a simulation system $\Sigma_S$ with $n$ objects implemented on some physical or formal machine $\Sigma_C$ (a computer). Assume that we need to know whether this computer is able to handle the integration of the system for the time interval $T_S$ (model time) within some pre-specified time interval $T_R$ (real time). In other words; $\Sigma_C$ needs to be updated $T_M$ time units model time, within $T_R$ time units real time.

Without any loss of generality we assume that $\Sigma_C$ is a sequential machine. Further assume that $s_i$ is the number of cpu cycles it takes to update the simplest object, $S_i$, and that $\tau_i$ is the (real) time of one cpu cycle. The minimal number of cpu cycles to update $\Sigma_S$ is therefor $ns_i$ and

$$\tau_{min} = ns_i\tau_i \tag{20}$$

therefore defines the minimal (real) time it takes to update the whole systems $\Sigma_S$ once.

$$R_{max} = \frac{1}{\tau_{min}} \tag{21}$$

therefor defines the maximal (real) rate by which $\Sigma_C$ can be updates. If we assume that the largest time increment (model time) allowed in each update of the system is $\Delta t$ then

$$N = \frac{T_M}{\Delta t} \tag{22}$$

defines the minimal number of system ($\Sigma_S$) updates to be done to complete the task.

The minimal number of updates left to be done at (real) time $t$ is therefore

$$Q_{min}(t) = ns_i(N - \frac{Q(t)}{n}) \tag{23}$$

where Q(t) is defined as the total number of updates performed by the machine at any given time $t$ (recall the definition of Q in the universal simulator in section (6)).

Thus, an on-line (optimistic) estimate of whether the desired integration can be accomplished within the specified time frame of $T_R$ time units, real time, can therefore be found through a comparison of the current, minimum update rate needed to complete the task

$$R_{needed}(t) = \int_0^t \frac{Q_{min}(\tau)}{T_R - \tau}d\tau \tag{24}$$

and the maximal, real time, rate by which $\Sigma_C$ can be updates (recall (21)).

If $R_{needed} > R_{max}$ at any time $t$ then the system $\Sigma_S$ is not *achievably* simulatable on $\Sigma_C$.

This follows directly of the definition of $R_{needed}$ and $R_{max}$.

## 8   Computability and Simulatability

Consider the iterated map

$$z(t+1) = z^2(t) + c, \qquad (25)$$

where $z$ and $c$ are complex numbers (see for instance [8]). It can be shown that for most $c$ the location of the closure of the unstable equilibrium set, the Julia set, for this mapping is *non-computable* where computation is defined over the real numbers [2] [3]. This means that it is *non-decidable* whether a given point in the complex plane is a member in the Julia set. Note that it is the observational function $O^2$, that decides membership in a set with certain properties, that is non-computable.

The above mapping is obviously simulatable and the Julia sets are also obviously emergent $S^2$ structures for this mapping. Thus, this example shows a simulatable system with non-computable emergent properties. The concepts of simulatability and computability are distinct for the purposes of the discussion here.

## 9   Conclusion

We have demonstrated that the foremost property of simulation is its ability to produce emergence. A simulation is an *emergence engine*. It is a representational mechanism that is distinguished by its capacity to generate relations that are not explicitly encoded. This ability enables us to study complicated dynamical properties which are otherwise intractable.

A system may be non-simulatable for certain (model, update) pairs pairs, but simulatable for other (model, update) pairs. Only when the update distributes over the ensemble of objects is a system simulatable.

We have defined and discussed the notion of a universal simulator and shown how the scheduling problem as well as the problem of load balancing naturally lives in this machine. Using the universal simulator we have shown that for any given set of model formulations of the interacting objects that constitute the system, together with a given update functional, it is possible in a finite number of operations to determine whether a system is simulatable or not. This is equivalent to a diagonal form of a corrosponding object update dependency Jacobian. This machine is also a useful practical device as an indicator for whether a system is (achievable) simulatable or not given a physical or formal machine (computer) where the simulation is implemented.

We have also demonstrated that a system may be simulatable, but have non-computable, emergent properties and thus the concepts of computability and simulatability are thus distinct for the purposes of the discussion here.

**Acknowledgments**

We would like to thank Nils Baas and Kai Nagel for constructive comments on earlier versions of the text as well as the many people at LANL and at SFI who have helped clarify our first ideas on the topic. It has been a long process.